\documentclass[aps,showpacs,twocolumn,superscriptaddress]{revtex4-1}
\usepackage{graphicx}
\usepackage{amsfonts}
\usepackage{amssymb}
\usepackage{amsmath}
\usepackage{natbib}
\usepackage{dcolumn}
\usepackage{bm}
\begin{document}
\title {Hyperon star in a modified quark meson coupling model}
\author{R.N. Mishra}
\author{H.S. Sahoo}
\affiliation{Department of Physics, Ravenshaw University, Cuttack-753 003, 
India}
\author{P.K. Panda}
\author{N. Barik}
\affiliation{Department of Physics, Utkal University, Bhubaneswar-751 004, 
India}
\author{T. Frederico}
\affiliation{ Instituto Tecn\'ologico de Aeron\'atica, DCTA,
12228-900 S\~ao Jos\'e dos Campos, SP, Brazil}
\begin{abstract}
We determine the equation of state (EOS) of nuclear matter with the 
inclusion of hyperons in a self-consistent manner by using a Modified 
Quark Meson Coupling Model (MQMC) where the confining interaction for 
quarks inside a baryon is represented by a phenomenological average 
potential in an equally mixed scalar-vector harmonic form. The hadron-hadron 
interaction in nuclear matter is then realized by introducing additional 
quark couplings to $\sigma$, $\omega$, and $\rho$ mesons through mean-field 
approximations. The effect of a nonlinear $\omega$-$\rho$ term on the 
equation of state is studied. The hyperon couplings are fixed from 
the optical potential values 
and the mass-radius curve is determined satisfying the maximum mass 
constraint of $2$~M$_{\odot}$ for neutron stars, as determined in recent 
measurements of the pulsar PSR J0348+0432. We also observe that there is 
no significant advantage of introducing the nonlinear $\omega$-$\rho$ term 
in the context of obtaining the star 
mass constraint in the present set of parametrizations.

\end{abstract}

\maketitle
\section{Introduction}
Over the last few decades intensive theoretical investigations 
are being pursued to understand the microscopic composition 
and properties of dense nuclear matter. It has been realized by 
now from such studies 
\cite{saakyan,pring,aakmal,jrstone,jrstone1,fweber,weise,bdlackey,debarati} 
that high density nuclear matter 
may consist not only of nucleons and leptons but also several 
exotic components such as hyperons, mesons as well as quark matter in 
different forms and phases. Hyperons in particular are expected to appear 
in the inner core of neutron stars at densities $2-3$ times the normal 
saturation density $\rho_0=0.15$ fm$^{-3}$. This is because at such high 
densities the nucleon chemical potential becomes large enough to facilitate 
the formation of hyperons to be energetically favorable by the inverse 
beta decay process of nucleons in the $\beta$-stable nuclear matter. As a 
consequence the Fermi pressure exerted by the baryons is reduced and the 
Equation of State (EOS) describing such dense matter in neutron stars with 
hyperon core becomes softer leading to the reduction of the maximum mass of 
the  star \cite{glend,glenmos,knorren,balberg,prakash,taurines}. However 
relativistic Hartree-Fock models \cite{miyatsu,miyatsu1}, 
relativistic mean field models \cite{bednarek,weissenborn} or quantum 
hadrodynamic model \cite{jiang} 
show relatively weaker effects on the EOS due to the presence of strange 
baryons in neutron star core.

Until recently the reliability requirement for any model EOS was only 
to predict a maximum neutron star mass $M_{max}$ compatible with the 
canonical value of $1.4-1.5$ M$_\odot$, since most of the precisely 
measured neutron star mass were clustered around these values only. 
This constraint was probably not stringent enough for which without 
any discrmination, most relativistic models even with the inclusion of 
hyperons \cite{glend,glenmos,knorren,balberg,prakash} have succeeded to 
this extent. But recent discovery of the 
unusually high mass of the millisecond pulsars PSR J1903+0327 
($1.66\pm0.021$~M$_{\odot}$) \cite{djchamp,freire,freire1}, PSR J1614-2230 
($1.97\pm0.04$~M$_{\odot}$) \cite{demorest} and PSR J0348+0432 
($2.01\pm0.04$~M$_{\odot}$) \cite{antoniadis} show that the neutron star mass 
distribution is much wider extending firmly up to $1.9-2.0$~M$_{\odot}$. 
Also there has been considerable progress in the measurement of the neutron 
star radii by reducing their uncertainties with a better understanding of the 
sources of systematic errors to estimate them in $10.1-11.1$ km range for a 
$1.5$~M$_\odot$ neutron star \cite{ozel}. Another study by Fortin {\it et al} 
\cite{fortin} has shown that the observational constraint on the maximum mass 
implies that the hyperonic stars with masses in the range $1-1.6$~M$_\odot$ 
must be larger than $13$ km due to a pre-hyperonic stiffening of EOS. It has 
been found by  Provid\^{e}ncia and Rabhi \cite{cprovindencia} that 
the radius of a hyperonic 
star of a given mass decreases linearly with the increase of the total hyperon 
content. These observations may serve to further constrain the EOS in 
achieving greater reliability.

Various studies have
established that the presence of hyperons in the neutron star core
leads to softening of the EOS and consequent reduction in the maximum mass of
the star. This has provided a 
challenge to develop an equation of state (EOS) stiff enough to give such 
high mass with the inclusion of hyperons. In fact most relativistic models 
obtain maximum star masses in the range $1.4-1.8 M_{\odot}$ with the inclusion 
of hyperons \cite{glenmos}. 
However there are some exceptional cases \cite{rikovska} where maximum mass 
of the hyperonic star have been realized in the range $1.8-2.1$~M$_\odot$.

In the present work, we have developed an EOS using a modified quark-meson 
coupling model (MQMC).  The MQMC model is based on  
confining relativistic independent quark potential model rather 
than a bag to describe the baryon structure in vacuum. In such a picture 
the quarks inside the baryon are considered to be independently confined 
by a phenomenologically average potential with an equally mixed scalar-vector 
harmonic form. Such a potential has  characteristically simplifying features 
in converting the independent quark Dirac equation into a Schr\"odinger 
like equation for the upper component of Dirac spinor which can be solved 
easily. The implications of such potential forms in the Dirac framework 
has been studied earlier \cite{barik,prd}. The baryon-baryon interactions 
are realized by 
making additional quark couplings to $\sigma$, $\omega$, and $\rho$ mesons 
through mean-field approximations, in an extension of previous works based 
on the MIT bag model \cite{guichon,qmc,frederico89}. 
The MQMC model has already been well tested in determining various bulk 
properties of symmetric and asymmetric nuclear matter \cite{rnm,hss}.
The relevant parameters of the interaction 
are obtained self-consistently by realizing the saturation properties 
such as binding energy and pressure.  
Here, we study the role of hyperons on the properties of neutron stars. 
In the present work we have also 
introduced an additional non-linear $\omega-\rho$ coupling to study its 
effect on the stiffening of EOS necessary for the purpose.

We include hyperons as a new degree of freedom in dense hadronic matter 
relevant for neutron stars. The interactions between nucleons and the baryons 
of the baryon octet in dense matter is studied and its effects on the 
mass of the neutron star is analysed. The nucleon-nucleon 
interaction is well known from nuclear properties. But the extrapolation of 
such interactions to densities beyond nuclear saturation density is a great 
problem. Most of the hyperon-nucleon interaction are known experimentally. 
This has inspired us to set the hyperon-nucleon interaction
potential at saturation density for the 
$\Lambda$, $\Sigma$ and $\Xi$ hyperons to $U_\Lambda = -28$ MeV, 
$U_\Sigma = 30$ MeV and $U_\Xi = -18$ MeV respectively. Because of the 
uncertainties in the measurement of the $\Xi$ hyperon potentials, we make a 
variation in the $U_\Xi$ and study the effects on the mass of the star. 
However, we do not include the hyperon-hyperon interactions which are 
experimentally least well known.

In this model we observe that the compressibility of the neutron star matter 
depends on the mass of the quark. The quark mass has been fixed at $150$ MeV 
 giving us a compressibility of $292$ MeV which lies within the range 
predicted from experimental GMR studies \cite{stonemos} and also from 
theoretical predictions of infinite nuclear matter model \cite{lsatpathy}. 
We also compare our 
results at two different quark masses of $m_q=150$ MeV and $m_q=80$ MeV.

The paper is organized as follows: In Sec. II, a brief outline
of the model describing the baryon structure in vacuum is discussed. 
The baryon mass is then 
realized by appropriately taking into account the center-of-mass correction, 
pionic correction, and gluonic correction in Sec. III. The EOS is then 
developed in Sec. IV. The results and discussions are made in Sec. V. 
We summarize our findings in Sec. IV.

\section{Modified quark meson coupling model}

The modified quark-meson coupling model has been extensively applied for the 
study of the bulk properties of both symmetric as well as 
asymmetric nuclear matter. Under such a model the nucleon-nucleon ($NN$) 
interaction was realized in a mean-field approach through the exchange of 
effective $(\sigma,\omega)$ mesonic fields coupling to the quarks inside 
the nucleon for the symmetric case \cite{rnm} and the additional 
iso-vector vector meson field ($\rho$) coupling to the quarks for the 
asymmetric case \cite{hss}. In our earlier work \cite{hss} this model was 
used to investigate the nature of the thermodynamic instabilities 
and the correlation of the symmetry energy with its slope. We now 
extend this model to investigate the role of nucleons and hyperons in 
neutron star matter under conditions of beta equilibrium and charge neutrality.

We begin by considering baryons as composed of three constituent quarks  
in a phenomenological flavor-independent confining potential, $U(r)$ in an 
equally mixed scalar and vector harmonic form inside the baryon \cite{rnm}, 
where 
\[
U(r)=\frac{1}{2}(1+\gamma^0)V(r),
\]
with 
\begin{equation}
V(r)=(ar^2+V_0),~~~~~ ~~~ a>0. 
\label{eq:1}
\end{equation}
Here $(a,~ V_0)$ are the potential parameters. The confining interaction  
provides the zeroth-order quark dynamics of the hadron.  In the medium, the 
quark field $\psi_q({\mathbf r})$ satisfies 
the Dirac equation
\begin{equation}
[\gamma^0~(\epsilon_q-V_\omega-
\frac{1}{2} \tau_{3q}V_\rho)-{\vec \gamma}.{\vec p}
-(m_q-V_\sigma)-U(r)]\psi_q(\vec r)=0
\end{equation}
where $V_\sigma=g_\sigma^q\sigma_0$, $V_\omega=g_\omega^q\omega_0$ and
$V_\rho=g_\rho^q b_{03}$. Here $\sigma_0$, $\omega_0$, and $b_{03}$ are the
classical meson fields, and 
$g_\sigma^q$, $g_\omega^q$, and $g_\rho^q$ are the quark couplings to  
the $\sigma$, $\omega$, and $\rho$ mesons, respectively. $m_q$ is the quark
mass and $\tau_{3q}$ is the third component of the Pauli matrices. We can now define
\begin{equation}
\epsilon^{\prime}_q= (\epsilon_q^*-V_0/2)~~~ 
\mbox{and}~~~ m^{\prime}_q=(m_q^*+V_0/2),
\label{eprim}
\end{equation}
where the effective quark energy, 
$\epsilon_q^*=\epsilon_q-V_\omega-\frac{1}{2}\tau_{3q} V_\rho$ and 
effective quark mass, $m_q^*=m_q-V_\sigma$. We now introduce $\lambda_q$ 
and $r_{0q}$ as
\begin{equation}
(\epsilon^{\prime}_q+m^{\prime}_q)=\lambda_q~~
~~\mbox{and}~~~~r_{0q}=(a\lambda_q)^{-\frac{1}{4}}.
\label{eq:8}
\end{equation}

The ground-state quark energy can be obtained from the eigenvalue condition
\begin{equation}
(\epsilon^{\prime}_q-m^{\prime}_q)\sqrt \frac{\lambda_q}{a}=3.
\label{eq:11}
\end{equation}
The solution of equation \eqref{eq:11} for the quark  energy
$\epsilon^*_q$ immediately leads to
the mass of baryon in the medium in zeroth order as
\begin{equation}
E_B^{*0}=\sum_q~\epsilon^*_q
\label{eq:12}
\end{equation}

\section{Effective mass of baryon}
We next consider the spurious center-of-mass 
correction $\epsilon_{c.m.}$, the pionic correction $\delta M_{B}^\pi$ 
for restoration of chiral symmetry, and the short-distance one-gluon 
exchange contribution $(\Delta E_B)_g$ to the zeroth-order baryon 
mass in the medium. 

Here, we extract the center of mass energy to first order in the 
difference between the fixed center and relative quark co-ordinate, 
using the method described by Guichon {\it et al.} \cite{guichon}. 
The centre of mass correction is given by:
\begin{equation}
e_{c.m.}=e_{c.m.}^{(1)}+e_{c.m.}^{(2)},
\end{equation}
where,

\begin{equation}
e_{c.m.}^{(1)}=\sum_{i=1}^3{\left[\frac{m_{q_i}}{\sum_{k=1}^3 m_{q_k}}\frac{6}
{r_{0q_i}^2(3\epsilon'_{q_i}+m'_{q_i})}\right]}
\end{equation}

\begin{widetext}
\begin{eqnarray}
e_{c.m.}^{(2)}&=&\frac{a}{2}\bigg[\frac{2}{\sum_k m_{q_k}}
\sum_im_i\langle r_i^2\rangle
+\frac{2}{\sum_k m_{q_k}}\sum_im_i\langle \gamma^0(i)r_i^2\rangle
-\frac{3}{(\sum_k m_{q_k})^2}\sum_im_i^2\langle r_i^2\rangle\nonumber\\
&-&\frac{1}{(\sum_k m_{q_k})^2}\sum_i\langle \gamma^0(1)m_i^2r_i^2\rangle-
\frac{1}{(\sum_k m_{q_k})^2}\sum_i\langle \gamma^0(2)m_i^2r_i^2\rangle-
\frac{1}{(\sum_k m_{q_k})^2}\sum_i\langle \gamma^0(3)m_i^2r_i^2\rangle\bigg]
\end{eqnarray}
\end{widetext}

 In the above, we have used for $i=(u,d,s)~~~$ and $k=(u,d,s)$ and the 
various quantities are defined as

\begin{equation}
\langle r_i^2\rangle = \frac{(11\epsilon_{qi}'+ m_{qi}')r^2_{0qi}}
{2(3\epsilon_{qi}'+ m_{qi}')}
\end{equation}

\begin{equation}
\langle\gamma^0(i)r_i^2\rangle=\frac{(\epsilon_{qi}'+ 11 m_{qi}')r^2_{0qi}}
{2(3\epsilon_{qi}'+ m_{qi}')}
\end{equation}

\begin{equation}
\langle\gamma^0(i)r_j^2\rangle_{i\neq j}=\frac{(\epsilon_{qi}'+ 3 m_{qi}')
\langle r^2_j\rangle}{3\epsilon_{qi}'+ m_{qi}'}
\end{equation}

The pseudo-vector nucleon pion coupling constant, $f_{NN\pi}$ can be obtained 
from Goldberg-Treiman relations by using the axial-vector coupling constant 
value $g_A$ in the model as
\begin{equation}
{\sqrt{4\pi}}\frac{f_{NN\pi}}{m_\pi}=\frac{g_A(N)}{2f_\pi},
\end{equation}
where
\begin{equation}
g_A(n\rightarrow p)= \frac{5}{9}\frac{(5\epsilon_u^{\prime}
+7m_u^{\prime})} {(3\epsilon_u^{\prime}+m_u^{\prime})}.
\end{equation}
The pionic corrections in the model for the nucleons become
\begin{equation}
\delta M_{N}^\pi=- \frac{171}{25}I_{\pi}f_{NN\pi}^2.
\end{equation}
Taking $w_k=(k^2+m_\pi^2)^{1/2}$
$I_{\pi}$ becomes
\begin{equation}
I_{\pi}=\frac{1}{\pi{m_{\pi}}^2}\int_{0}^{\infty}dk. 
\frac{k^4u^2(k)}{w_k^2},
\end{equation}
with the axial vector nucleon form factor given as
\begin{equation}
u(k)=\Big[1-\frac{3}{2} \frac{k^2}{{\lambda}_q(5\epsilon_q^{\prime}+
7m_q^{\prime})}\Big]e^{-k^2r_0^2/4} \ .
\end{equation}
The pionic correction for $\Sigma^{0}$ and $\Lambda^{0}$ become
\begin{equation}
\delta M_{\Sigma^{0}}^{\pi}=-{\frac{12}{5}}f_{NN\pi}^2I_{\pi},
\end{equation}
\begin{equation}
\delta M_{\Lambda^{0}}^{\pi}=-{\frac{108}{25}}f_{NN\pi}^2I_{\pi}.
\end{equation}
Similarly the pionic correction for $\Sigma^{-}$ and $\Sigma^{+}$ is
\begin{equation}
\delta M_{\Sigma^{+},\Sigma^{-}}^{\pi}=-{\frac{12}{5}}f_{NN\pi}^2I_{\pi}.
\end{equation}
The pionic correction for $\Xi^{0}$ and $\Xi^{-}$ is
\begin{equation}
\delta M_{\Xi^{-},\Xi^{0}}^{\pi}=-{\frac{27}{25}}f_{NN\pi}^2I_{\pi}.
\end{equation}

The one-gluon exchange interaction is provided by the interaction Lagrangian
density
\begin{equation}
{\cal L}_I^g=\sum J^{\mu a}_i(x)A_\mu^a(x) \ ,
\end{equation}
where $A_\mu^a(x)$ are the octet gluon vector-fields and $J^{\mu a}_i(x)$ is 
the $i$-th quark color current. The gluonic correction can be separated in two
pieces, namely, one from the color electric field ($E^a_i$) and another 
from the magnetic field ($B^a_i$)
generated by the $i$-th quark color current density
\begin{equation}
J^{\mu a}_i(x)=g_c\bar\psi_q(x)\gamma^\mu\lambda_i^a\psi_q(x) \ ,
\end{equation}
with $\lambda_i^a$ being the usual Gell-Mann $SU(3)$ matrices and
$\alpha_c=g_c^2/4\pi$. The contribution to the mass 
can be written as a sum of color electric and color magnetic part as
\begin{equation}
(\Delta E_B)_g=(\Delta E_B)_g^E+(\Delta E_B)_g^M \ ,
\end{equation}
where
\begin{eqnarray}
(\Delta E_B)_g^E &=&\frac{1}{8\pi}\sum_{i,j}\sum_{a=1}^8\int\frac{d^3r_id^3r_j}
{|r_i-r_j|}\nonumber\\
&\times&\langle B |J^{0 a}_i(r_i)J^{0 a}_j(r_j)|B\rangle \ ,
\end{eqnarray}
and
\begin{eqnarray}
(\Delta E_B)_g^M&=&-\frac{1}{8\pi}\sum_{i,j}\sum_{a=1}^8\int\frac{d^3r_id^3r_j}
{|r_i-r_j|}\nonumber\\
&\times& \langle B |\vec J^a_i(r_i)\vec J^a_j(r_j)|B\rangle \ .\\ \nonumber
\end{eqnarray}

Finally, taking into account the specific quark flavor and spin configurations
in the ground state baryons and using the relations
$\langle\sum_a(\lambda_i^a)^2\rangle =16/3$ and
$\langle\sum_a(\lambda_i^a\lambda_j^a)\rangle_{i\ne j}=-8/3$ for
baryons, one can write the energy correction due to
color electric contribution, as
\begin{equation}
(\Delta E_B)_g^E={\alpha_c}(b_{uu}I_{uu}^E+b_{us}I_{us}^E+b_{ss}I_{ss}^E) \ ,   
\label{enge}
\end{equation}
and due to color magnetic contributions, as
\begin{equation}
(\Delta E_B)_g^M={\alpha_c}(a_{uu}I_{uu}^M+a_{us}I_{us}^M+a_{ss}I_{ss}^M) \ ,  
\label{engm}
\end{equation}
where $a_{ij}$ and $b_{ij}$ are the numerical coefficients depending on each
baryon and are given in Table \ref{table0}. In the above, we have

\begin{table}[t]
\renewcommand{\arraystretch}{1.4}
\setlength\tabcolsep{3pt}
\begin{tabular}{|c|c|c|c|c|c|c|}
\hline
Baryon     & $a_{uu}$ & $a_{us}$ & $a_{ss}$ & $b_{uu}$ & $b_{us}$ & $b_{ss}$\\
\hline
$N$        & -3  &  0  & 0 & 0 &  0 & 0\\
$\Lambda$  & -3  &  0  & 0 & 1 & -2 & 1\\
$\Sigma$   &  1  & -4  & 0 & 1 & -2 & 1\\
$\Xi$      &  0  & -4  & 1 & 1 & -2 & 1\\
\hline
\end{tabular}
\caption{\label{table0}The coefficients $a_{ij}$ and $b_{ij}$ used in the 
calculation of the color-electric and and color-magnetic energy 
contributions due to one-gluon exchange.}
\end{table}

\begin{eqnarray}
I_{ij}^{E}=\frac{16}{3{\sqrt \pi}}\frac{1}{R_{ij}}\Bigl[1-
\frac{\alpha_i+\alpha_j}{R_{ij}^2}+\frac{3\alpha_i\alpha_j}{R_{ij}^4}
\Bigl]
\nonumber\\
I_{ij}^{M}=\frac{256}{9{\sqrt \pi}}\frac{1}{R_{ij}^3}\frac{1}{(3\epsilon_i^{'}
+m_{i}^{'})}\frac{1}{(3\epsilon_j^{'}+m_{j}^{'})} \ ,
\end{eqnarray}
where
\begin{eqnarray}
R_{ij}^{2}&=&3\Bigl[\frac{1}{({\epsilon_i^{'}}^2-{m_i^{'}}^2)}+
\frac{1}{({\epsilon_j^{'}}^2-{m_j^{'}}^2)}\Bigl]
\nonumber\\
\alpha_i&=&\frac{1}{ (\epsilon_i^{'}+m_i^{'})(3\epsilon_i^{'}+m_{i}^{'})} \ .
\end{eqnarray}
The color electric contributions to the bare mass for nucleon 
$(\Delta E_N)_g^{E} = 0$. Therefore the one-gluon contribution for nucleon 
becomes
\begin{equation}
(\Delta E_N)_g^{M}=-{\frac{256\alpha_c}{3\sqrt{\pi}}}
\Big [{\frac{1}{(3{\epsilon}_u^{\prime}+m_u^{\prime})^2R_{uu}^3}}\Big ]\\
\end{equation}

The one-gluon contribution for $\Sigma^{+} ,\Sigma^{-}$ becomes
\begin{eqnarray}
(\Delta E_{\Sigma^{+} ,\Sigma^{-}})_g^{E}&=&{\alpha_c}{\frac{16}{3\sqrt{\pi}}}
\Bigg [{\frac{1}{R_{uu}}}\left(1-{\frac{2\alpha_u}{R_{uu}^2}}-
{\frac{3\alpha_u^2}{R_{uu}^4}}\right) \nonumber \\   
&-&{\frac{2}{R_{us}}}\left(1-{\frac{\alpha_u+\alpha_s}{R_{us}^2}} +
{\frac{3\alpha_u\alpha_s}{R_{us}^4}}\right) \nonumber\\
&+&{\frac{1}{R_{ss}}}\left(1-{\frac{2\alpha_s}{R_{ss}^2}}+
{\frac{3\alpha_s^2}{R_{ss}^4}}\right)\Bigg ]
\end{eqnarray}

\begin{eqnarray}
(\Delta E_{\Sigma^{+} ,\Sigma^{-}})_g^{M}&=&{\frac{256\alpha_c}{9\sqrt{\pi}}}
\Bigg [{\frac{1}{(3{\epsilon}_u^{\prime}+m_u^{\prime})^2R_{uu}^3}}\nonumber\\ 
&-&{\frac{4}
{R_{us}^3(3{\epsilon}_u^{\prime}+m_u^{\prime})(3{\epsilon}_s^{\prime}+
m_s^{\prime})}}\Bigg ]
\end{eqnarray}
\begin{equation}
(\Delta E_{\Sigma^{+} ,\Sigma^{-}})_g=(\Delta E_{\Sigma^{+} ,
\Sigma^{-}})_g^{E}+(\Delta E_{\Sigma^{+} ,\Sigma^{-}})_g^M
\end{equation}
The gluonic correction for $\Sigma^{0}$ is
\begin{eqnarray}
(\Delta E_{\Sigma^{0}})_g^{E}&=&{\alpha_c}{\frac{16}{3\sqrt{\pi}}}
\Bigg [{\frac{1}{R_{uu}}}\left(1-{\frac{2\alpha_u}{R_{uu}^2}}-
{\frac{3\alpha_u^2}{R_{uu}^4}}\right)  \nonumber \\   
&-&{\frac{2}{R_{us}}}\left(1-{\frac{\alpha_u+\alpha_s}{R_{us}^2}} +
{\frac{3\alpha_u\alpha_s}{R_{us}^4}}\right) \nonumber\\
&+&{\frac{1}{R_{ss}}}\left(1-{\frac{2\alpha_s}{R_{ss}^2}}+
{\frac{3\alpha_s^2}{R_{ss}^4}}\right)\Bigg ]
\end{eqnarray}

\begin{eqnarray}
(\Delta E_{\Sigma^{0}})_g^{M}&=&{\frac{256\alpha_c}{9\sqrt{\pi}}}
\Bigg [{\frac{1}{(3{\epsilon}_u^{\prime} +m_u^{\prime})^2R_{uu}^3}}\nonumber\\ 
&-&{\frac{4}
{R_{us}^3(3{\epsilon}_u^{\prime}+m_u^{\prime})(3{\epsilon}_s^{\prime}+
m_s^{\prime})}}\Bigg ]
\end{eqnarray}
\begin{equation}
(\Delta E_{\Sigma^{0}})_g=(\Delta E_{\Sigma^{0}})_g^{E}
+(\Delta E_{\Sigma^{0}})_g^M
\end{equation}

The gluonic correction for $\Lambda$ is
\begin{equation}
(\Delta E_{\Sigma^{0}})_g^{E}=(\Delta E_{\Lambda})_g^{E}
\end{equation}
The color magnetic contribution is different
\begin{equation}
(\Delta E_{\Lambda})_g^{M}=-{\frac{256\alpha_c}{3\sqrt{\pi}}}\Bigg [
{\frac{1}{(3{\epsilon}_u^{\prime} +m_u^{\prime})^2R_{uu}^3}}\Bigg ]
\end{equation}
\begin{equation}
(\Delta E_{\Lambda})_g=(\Delta E_{\Lambda})_g^{E}
+(\Delta E_{\Lambda})_g^M
\end{equation}

The color electric contributions for $\Xi^{-}$ and $\Xi^{0}$ are same as
that of $\Sigma^{0}$ or $\Lambda^{0}$ but the color magnetic contributions
to the correction of masses of baryon  are different:
\begin{eqnarray}
(\Delta E_{\Xi^{-} ,\Xi^{0}})_g^{M}&=&{\frac{256\alpha_c}{9\sqrt{\pi}}}
\Bigg [{\frac{1}{(3{\epsilon}_s^{\prime}+m_s^{\prime})^2R_{ss}^3}}\nonumber\\ 
&-&{\frac{4}
{R_{us}^3(3{\epsilon}_u^{\prime}+m_u^{\prime})(3{\epsilon}_s^{\prime}+
m_s^{\prime})}}\Bigg ]
\end{eqnarray}

Finally, the gluonic correction for $\Xi^{-}$ and $\Xi^{0}$ is given by:
\begin{equation}
(\Delta E_{\Xi^{-} ,\Xi^{0}})_g=(\Delta E_{\Xi^{-} ,\Xi^{0}})_g^{E}+
(\Delta E_{\Xi^{-} ,\Xi^{0}})_g^M
\end{equation}

Treating all energy corrections independently, the 
mass of the baryon in the medium becomes 
\begin{equation}
M_B^*=E_B^{*0}-\epsilon_{c.m.}+\delta M_B^\pi+(\Delta E_B)^E_g+
(\Delta E_B)^M_g.
\label{mass}
\end{equation}

\section{The Equation of state}
The total energy density and pressure at a 
particular baryon density, encompassing all the members of the baryon octet, 
for the nuclear matter in $\beta$-equilibrium can be found as
\begin{subequations}
\begin{eqnarray}
\label{engd}
{\cal E} &=&\frac{1}{2}m_\sigma^2 \sigma_0^2+\frac{1}{2}m_\omega^2 \omega^2_0
+\frac{1}{2}m_\rho^2 b^2_{03} + 3g_\omega^2g_\rho^2\Lambda_\nu b_{03}^2\omega_0^2
\nonumber\\ &+&\frac{\gamma}{2\pi^2}\sum_{B}\int ^{k_{f,B}} 
[k^2+{M_B^*}^2]^{1/2}k^2dk
\nonumber\\
&+&\sum_{l}\frac{1}{\pi^2}\int_0^{k_l}[k^2+m_l^2]^{1/2}k^2dk,\\
P&=&-~\frac{1}{2}m_\sigma^2 \sigma_0^2+\frac{1}{2}m_\omega^2 \omega^2_0+
\frac{1}{2}m_\rho^2 b_{03}^2+ g_\omega^2g_\rho^2\Lambda_\nu b_{03}^2\omega_0^2
\nonumber\\
&+&\frac{\gamma}{6\pi^2}\sum_{B}\int ^{k_{f,B}} \frac{k^4~ dk}
{[k^2+{M_B^*}^2]^{1/2}} \nonumber\\
&+& \frac{1}{3}\sum_{l}\frac{1}{\pi^2}\int_0^{k_l}
\frac{k^4dk}{[k^2+m_l^2]^{1/2}},
\end{eqnarray} 
\end{subequations}
where $\gamma=2$ is the spin degeneracy factor for nuclear matter,
$B=N,\Lambda,~\Sigma^{\pm},~\Sigma^0,~\Xi^-,~\Xi^0$ and $l=e,\mu$.
In the above expression for the energy density and pressure, a 
nonlinear $\omega-\rho$ coupling term is introduced with coupling
coefficient, $\Lambda_\nu$ \cite{horowitz01}.

Another important quantity for the study of nuclear matter is the symmetry 
energy, which is defined as
\begin{equation}
{\cal E}_{sym}(\rho_B)=\frac{k^2}{6E_N^{*2}}
+\frac{g_\rho^2} {8m_\rho^{2}}\rho_B
\label{engs}
\end{equation}
where $E_N^*=\sqrt{k^2+M_N^{*2}}$, the index $N=n,p$ for neutrons and protons. 
The slope of the 
symmetry energy $L$ is then obtained as,
\begin{equation}
L=3\rho_0\dfrac{\partial {\cal E}_{sym}(\rho_B)}{\partial \rho_B}
\Bigg|_{\rho_B=\rho_0}
\end{equation}
For obtaining a constraint on the quark mass we use the value of 
compressibility given by,
\begin{equation}
K=9\left[\frac{dP}{d\rho_B}\right]_{\rho_B=\rho_0}
\end{equation}

The chemical potentials, necessary to define the $\beta-$ 
equilibrium conditions, are given by
\begin{equation}
\mu_B=\sqrt{k_B^2+{M_B^*}^2}+g_\omega\omega_0+g_\rho\tau_{3B}b_{03}
\end{equation}
where $\tau_{3B}$ is the isopsin projection of the baryon B.

The lepton Fermi momenta are the positive real solutions of
$(k_e^2 + m_e^2)^{1/2} =  \mu_e$ and
$(k_\mu^2 + m_\mu^2)^{1/2} = \mu_\mu$. The equilibrium composition
of the star is obtained by solving the equations of motion of meson fields in 
conjunction with the charge neutrality condition, given in equation 
(\ref{neutral}),  
at a given total baryonic density $\rho = \sum_B \gamma k_B^3/(6\pi^2)$. 
The effective masses of the baryons are
obtained self-consistently in this model.

Since we consider the octet baryons, the presence of strange baryons 
in the matter plays a significant role. We define the strangeness 
fraction as 
\begin{equation}
f_s=\frac{1}{3}\frac{\sum_i |s_i|\rho_i}{\rho}.
\end{equation}
Here $s_i$ refers to the strangeness number of baryon $i$ and  
$\rho_i$ is defined as $\rho_i=\gamma k_{Bi}^3/(6\pi^2)$.

For stars in which the strongly interacting particles are baryons, the
composition is determined by the requirements of charge neutrality
and $\beta$-equilibrium conditions under the weak processes
$B_1 \to B_2 + l + {\overline \nu}_l$ and $B_2 + l \to B_1 + \nu_l$.
After deleptonization, the charge neutrality condition yields
\begin{equation}
q_{\rm tot} = \sum_B q_B \frac{\gamma k_B^3}{6\pi^2}
+ \sum_{l=e,\mu} q_l \frac{k_l^3}{3\pi^2}  = 0 ~,
\label{neutral}
\end{equation}
where $q_B$ corresponds to the electric charge of baryon species $B$
and $q_l$ corresponds to the electric charge of lepton species $l$. Since
the time scale of a star is effectively infinite compared to the weak
interaction time scale, weak interaction violates strangeness conservation.
The strangeness quantum number is therefore not conserved
in a star and the net strangeness is determined by the condition of
$\beta$-equilibrium which for baryon $B$ is then given by
$\mu_B = b_B\mu_n - q_B\mu_e$, where $\mu_B$ is the chemical potential
of baryon $B$ and $b_B$ its baryon number. Thus the chemical potential of any
baryon can be obtained from the two independent chemical potentials $\mu_n$
and $\mu_e$ of neutron and electron respectively.

The hyperon couplings are not
relevant to the ground state properties of nuclear matter, but information
about them can be available from the levels in $\Lambda$ hypernuclei
\cite{chrien}.
$$g_{\sigma B}=x_{\sigma B}~ g_{\sigma N},~~g_{\omega B}=x_{\omega B}~ 
g_{\omega N}, ~~g_{\rho B}=x_{\rho B}~ g_{\rho N}$$
and $x_{\sigma B}$, $x_{\omega B}$ and $x_{\rho B}$ are equal to $1$ for the
nucleons and acquire different values in different parameterisations for the
other baryons. We note that the $s$-quark is unaffected by the sigma and omega
mesons i.e. $g_\sigma^s=g_\omega^s=0$ .

The vector mean-fields $\omega_0$ and $b_{03}$ are determined through
\begin{equation}
\omega_0=\frac{g_\omega}{{m_\omega^*}^2} \sum_B x_{\omega B}\rho_B~~~~~
b_{03}=\frac{g_\rho}{{2m_\rho^*}^2} \sum_B x_{\rho B}\tau_{3B}\rho_B,
\label{omg}
\end{equation}
where ${m_\omega^*}^2=m_\omega^2+2\Lambda_{\nu}g_\rho^2g_\omega^2b_{03}^2$, 
${m_\rho^*}^2=m_\rho^2+2\Lambda_{\nu}g_\rho^2g_\omega^2\omega_0^2$, 
$g_\omega=3 g_\omega^q$ and $g_\rho= g_\rho^q$.
Finally, the scalar mean-field $\sigma_0$ is fixed by
\begin{equation}
\frac{\partial {\cal E }}{\partial \sigma_0}=0.
\label{sig}
\end{equation}
The iso-scalar scalar and iso-scalar vector couplings $g_\sigma^q$ and $g_\omega$
are fitted to the saturation density and binding energy for nuclear
matter. The iso-vector vector coupling $g_\rho$ is set by fixing 
the symmetry energy at $J=32.0$ MeV. 
For a given baryon density, $\omega_0$, $b_{03}$, and $\sigma_0$ are
calculated from equations \eqref{omg} and \eqref{sig}, respectively. 

Following the determination of the EOS the relation between the mass and 
radius of a star with its central density can be obtained by integrating 
the Tolman-Oppenheimer-Volkoff (TOV) equations \cite{tov} given by,
\begin{equation}
\frac{dP}{dr}=-\frac{G}{r}\frac{\left[{\cal E}+P\right ]\left[M+
4\pi r^3 P\right ]}{(r-2 GM)},
\label{tov1}
\end{equation}
\begin{equation}
\frac{dM}{dr}= 4\pi r^2 {\cal E} ,
\label{tov2}
\end{equation}
with $G$ as the gravitational constant and $M(r)$ as the enclosed gravitational
mass. We have used $c=1$. Given an EOS, these equations can be integrated 
from the origin as an initial value problem for a given choice of the 
central energy density, $(\varepsilon_0)$.
Of particular importance is the maximum mass obtained from and the 
solution of the TOV equations. 
The value of $r~(=R)$, where the pressure vanishes defines the
surface of the star. 

\section{Results and Discussion}
Our MQMC model has two potential parameters, $a$ and  $V_0$ and we obtain them 
by fitting the nucleon mass $M_N=939$ MeV and charge radius of the 
proton $\langle r_N\rangle=0.87$ fm in free space. Keeping the value of the 
potential parameter $a$ same as that for nucleons, we obtain $V_0$ for the 
$\Lambda$, $\Sigma$ and $\Xi$ baryons by fitting their respective masses 
to $M_{\Lambda}=1115.6$ MeV, $M_{\Sigma}=1193.1$ MeV and $M_{\Xi}=1321.3$ MeV. 
The set of potential parameters for the baryons along with their respective 
energy corrections at zero density are given in Table \ref{table1}. 
\begin{table*}[ht]
\renewcommand{\arraystretch}{1.4}
\setlength\tabcolsep{3pt}
\begin{tabular}{|c|c|c|c|c|c|c|c|c|c|}
\hline
Baryon& $M_B$ &\multicolumn{4}{c|}{$m_q=80$ MeV}&\multicolumn{4}{c|}
{$m_q=150$ MeV}\\ 
\cline{3-10}
       & (MeV) & $V_0$ (MeV)& e$_{c.m.}$ (MeV)&$\delta_B^\pi$ (MeV)&
($\Delta$E$_B$)$_g$ (MeV)&$V_0$ (MeV)& e$_{c.m.}$ (MeV)&$\delta_B^\pi$ (MeV)&
($\Delta$E$_B$)$_g$ (MeV)\\
\hline
N      &  939  &82.93 &357.92 &-72.52 &-68.69 &44.05 &331.84 &-86.96 &-59.02 \\  
$\Lambda$ &1115.6&87.03 &317.80 &-46.43 &-65.34 &50.06 &310.39 &-55.82 &-56.13\\
$\Sigma$ &1193.1&105.27&316.16 &-27.36 &-52.87 &66.44 &308.84 &-32.38 &-45.00\\  
$\Xi$    &1321.3&114.43&319.79 &-12.67 &-57.65 &66.82 &302.17 &-14.58 &-49.64\\ 
\hline
\end{tabular}
\caption{\label{table1}The potential parameter $V_0$ obtained for the 
quark mass $m_u=m_d=80$ MeV, $m_s=230$ MeV with  $a=0.81006$~fm$^{-3}$ and 
the quark mass $m_u=m_d=150$ MeV, $m_s=300$ MeV with  $a=0.69655$~fm$^{-3}$. 
Also shown are the contribution of the center of mass correction, pionic 
correction and gluonic correction to the baryon mass in free space.}
\end{table*} 
The quark meson couplings $g_\sigma^q$, $g_\omega=3g_\omega^q$, 
and $g_\rho=g_\rho^q$ are fitted self-consistently for the nucleons to obtain 
the correct saturation properties of nuclear matter binding energy, 
$E_{B.E.}\equiv B_0={\cal E}/\rho_B-M_N=-15.7$ MeV, pressure, $P=0$, and 
symmetry energy $J=32.0$ MeV at 
$\rho_B=\rho_0=0.15$ fm$^{-3}$. 

Table \ref{table01} shows 
the contribution to the spurious center-of-mass correction, the 
pionic correction and the gluonic correction to obtain the effective mass 
of the baryon. It is interesting to note that as the mass of the quark 
increases from $80$ MeV to $150$ MeV, the magnitude of the pionic correction 
increases whereas that due gluonic correction decreases for all baryon species.

We have taken the standard values for the 
meson masses; namely, $m_\sigma=550$ MeV, $m_\omega=783$ MeV and 
$m_\rho=763$ MeV. The values of the quark meson couplings, 
$g_\sigma^q$, $g_\omega$, and $g_\rho$ at quark masses $80$ MeV and 
$150$ MeV are given in Table \ref{table2}.
\begin{table*}[ht]
\renewcommand{\arraystretch}{1.4}
\setlength\tabcolsep{3pt}
\begin{tabular}{|c|c|c|c|c|c|c|c|c|}
\hline
Baryon& \multicolumn{4}{c|}{$m_q=80$ MeV} &\multicolumn
{4}{c|}{$m_q=150$ MeV}\\
\cline{2-9}
 & $M_B^*$ (MeV)& e$_{c.m.}$ (MeV)& $\delta_B^\pi$ (MeV)& 
($\Delta$E$_B$)$_g$ (MeV)& $M_B^*$ (MeV)&e$_{c.m.}$ (MeV)& 
$\delta_B^\pi$ (MeV)& ($\Delta$E$_B$)$_g$ (MeV)\\
\hline
N         &834.03   &364.64 &-35.40 &-77.84 &797.29  &344.38 &-46.13 &-69.39\\  
$\Lambda$ &1039.49  &326.18 &-46.45 &-48.29 &1018.10 &322.28 &-57.05 &-33.28\\
$\Sigma$  &1109.39  &324.65 &-27.47 &-40.59 &1087.53 &320.77 &-33.17 &-28.47\\  
$\Xi$     &1289.59  &322.88 &-12.74 &-41.86 &1282.12 &307.00 &-14.94 &-28.92\\  
\hline
\end{tabular}
\caption{\label{table01} The contribution of the center of mass correction, 
pionic correction and gluonic correction to the effective mass $M_B^*$ of the 
baryon at saturation density for quark mass $m_u=m_d=80$ MeV, $m_s=230$ MeV 
and $m_u=m_d=150$ MeV, $m_s=300$ MeV.} 
\end{table*} 

\begin{table*}[ht]
\centering
\renewcommand{\arraystretch}{1.4}
\setlength\tabcolsep{3pt}
\begin{tabular}{|c|c|c|c|c|c|c|c|c|c|c|c|c|c|c|}
\hline
$m_q$ & $g^q_\sigma$& $g_\omega$&\multicolumn{3}{c|}{ $g_\rho$}&$\sigma_0$ 
& $M_N^*/M_N$& K & \multicolumn{3}{c|}{L (MeV)}\\
\cline{4-6}\cline{10-12}
(MeV)& & &$\Lambda_\nu=0$&$\Lambda_\nu=0.05$&$\Lambda_\nu=0.1$ &(MeV)& 
& (MeV)& $\Lambda_\nu=0$&$\Lambda_\nu=0.05$&$\Lambda_\nu=0.1$\\
\hline
80    &4.89039  &5.17979 &8.92265&9.0790&9.2440 &13.34 &0.88 &246 &85.44
&87.53&89.77\\
\hline
150   &4.39952  &6.74299 &8.79976&9.2522&9.7825 &14.44 &0.87 &292 &86.39
&92.45&99.95\\
\hline
\end{tabular}
\caption{\label{table2}Parameters for nuclear matter. They are determined
from the binding energy per nucleon, $E_{B.E}=B_0 \equiv{\cal E} /\rho_B - M_N 
= -15.7$~MeV and pressure, $P=0$ at saturation density
$\rho_B=\rho_0=0.15$~fm$^{-3}$. Also shown are the values of the nuclear 
matter incompressibility $K$ and the slope of the symmetry energy $L$ 
for the quark masses $m_q = 80$ MeV and $m_q=150$ MeV.}
\end{table*}
By changing the value of the 
$\omega$-$\rho$ coupling term $\Lambda_\nu$ there is a change in the 
value of $g_\rho$. For $\Lambda_\nu=0.05$ and $0.1$ we obtain
the value of $g_\rho$ to be $9.25223$ and $9.78255$ respectively.

Incompressibility $K$ of symmetric nuclear matter as well as the slope of the 
symmetry energy $L$ provide important constraints to the properties of 
nuclear matter. 
In the present work, we determine the value of the compression modulus $K$ at 
quark masses $80$ MeV and $150$ MeV which comes out to be $246$ MeV and $292$ 
MeV respectively. From various experimental giant monopole resonance (GMR) 
studies \cite{stonemos} and microscopic calculations of the GMR energies 
\cite{khan} the value of $K$ is predicted to lie in the range $250<K<325$ MeV 
and $230 \pm 40$ MeV respectively. The slope of the nuclear symmetry energy $L$ 
in the present work is calculated to be $85.44$ MeV and $86.39$ MeV for 
quark masses $80$ MeV and $150$ MeV, which agrees well with the 
value $88 \pm 25$ extracted from isospin sensitive observables in 
heavy-ion reactions \cite{xuli}. By increasing the value of $\Lambda_\nu$ 
the value of $L$ increases to $L=92.45$ for $\Lambda_\nu=0.05$ and 
$L=99.95$ for $\Lambda_\nu=0.1$.
\begin{figure}
\includegraphics[width=8.cm,angle=0]{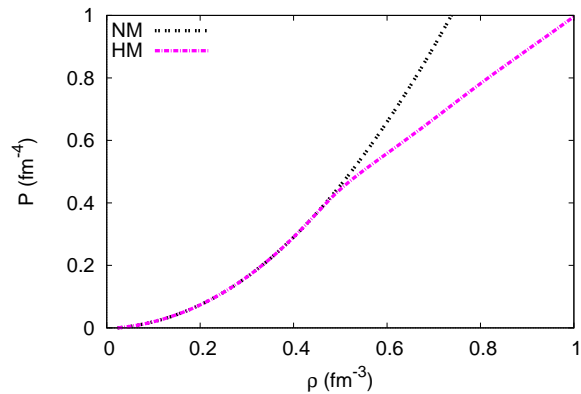}
\caption{(Color online) The pressure at various densities for nuclear 
matter (NM) and hyperon matter (HM).}
\label{fig1}
\end{figure}
\begin{figure}
\includegraphics[width=8.cm,angle=0]{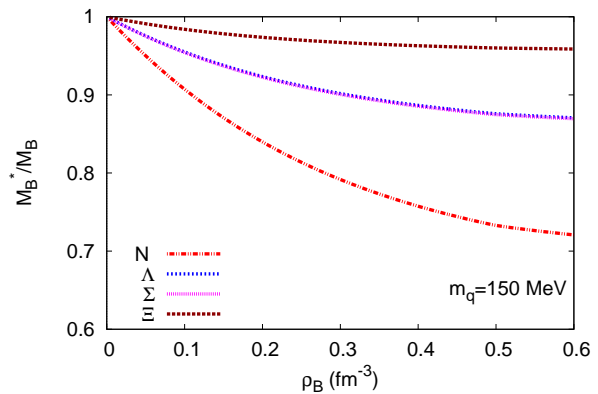}
\caption{(Color online) Effective mass of baryon at quark mass $m_q=150$ MeV.}
\label{fig2}
\end{figure}
\begin{figure}
\includegraphics[width=8.cm,angle=0]{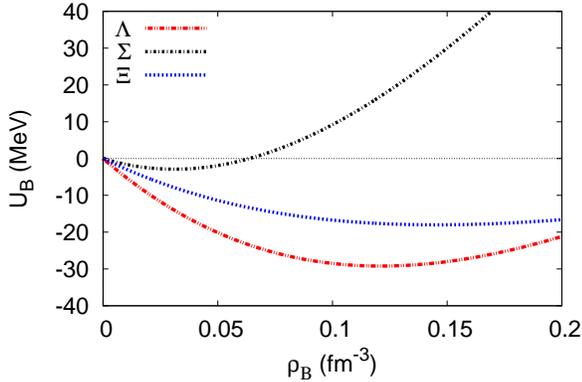}
\caption{(Color online) Hyperon ($\Lambda$, $\Sigma$, $\Xi$) potentials as a function 
of density.}
\label{fig3}
\end{figure}
The couplings of the hyperons to the $\sigma$-meson need not be fixed 
since we determine the effective masses of the hyperons self-consistently. 
The hyperon couplings to the $\omega$-meson 
are fixed by determining $x_{{\omega}B}$. 
The value of $x_{{\omega}B}$ is obtained from the hyperon potentials in nuclear 
matter, $U_B=-(M_B-M_B^*)+x_{{\omega}B}g_{\omega}\omega_0$ for $B={\Lambda},
{\Sigma}$ and $\Xi$ as $-28$ MeV, $30$ MeV and $-18$ MeV respectively. 
For the quark masses $80$ MeV and $150$ MeV the corresponding values 
for $x_{{\omega}B}$ are given in 
Table \ref{table3}. 
\begin{table*}[ht]
\renewcommand{\arraystretch}{1.4}
\setlength\tabcolsep{3pt}
\begin{tabular}{|c|c|c|c|c|c|}
\hline
$m_q$ & $x_{{\omega}\Lambda}$& $x_{{\omega}\Sigma}$&
\multicolumn{3}{c|}{$x_{{\omega}\Xi}$}\\ 
\cline{2-6}
(MeV)& $U_\Lambda=-28$ MeV& $U_\Sigma=30$ MeV& $U_\Xi=-18$ MeV& 
$U_\Xi=-10$MeV& $U_\Xi=0$ MeV\\
\hline
80      &0.95375    & 2.25435  & 0.27168 &0.43029 & 0.62857\\
\hline
150     &0.81309  & 1.58607 & 0.24769 &0.34129 & 0.45829 \\
\hline
\end{tabular}
\caption{\label{table3} $x_{{\omega}B}$ determined by fixing the 
 potentials for the hyperons.}
\end{table*}
The value of $x_{{\rho}B}=1$ is fixed for all baryons. 

The $\Lambda$ hyperon potential has been chosen from the measured single  
particle  levels of $\Lambda$ hypernuclei from mass numbers $A = 3$ to $209$ 
\cite{millener} of the binding of $\Lambda$ to symmetric nuclear matter. 
Studies of $\Sigma$ nuclear interaction \cite{mares,bart} from the analysis 
of $\Sigma^-$ atomic data indicate a repulsive isoscalar potential in the 
interior of nuclei. 
However, measurements of the $\Xi$ hyperon potential exhibit uncertainties. 
Measurements of the final state interaction of $\Xi$ hyperons produced in 
($K^-,K^+$) reaction on $^{12}C$ in E224 experiment at KEK \cite{fukuda} and 
E885 experiment at AGS \cite{khaustov} indicate a shallow attractive 
potential $U_\Xi \sim -16$ MeV and $U_\Xi \sim-14$ or less respectively. 
Hence, to study the effect of the coupling 
to the cascade we show the results at $U_\Xi = -10$ MeV and $U_\Xi = 0$ MeV 
in addition to $U_\Xi = -18$ MeV. For $U_\Xi = -10$ MeV 
$x_{{\omega}\Xi} = 0.43029$ at $m_q=80$ MeV and 
$x_{{\omega}\Xi} = 0.34129$ at $m_q=150$ MeV. 
For $U_\Xi = 0$ MeV $x_{{\omega}\Xi} = 0.62857$ at $m_q=80$ MeV and
$x_{{\omega}\Xi} = 0.45829$ at $m_q=150$ MeV.
\begin{figure}
\includegraphics[width=8.cm,angle=0]{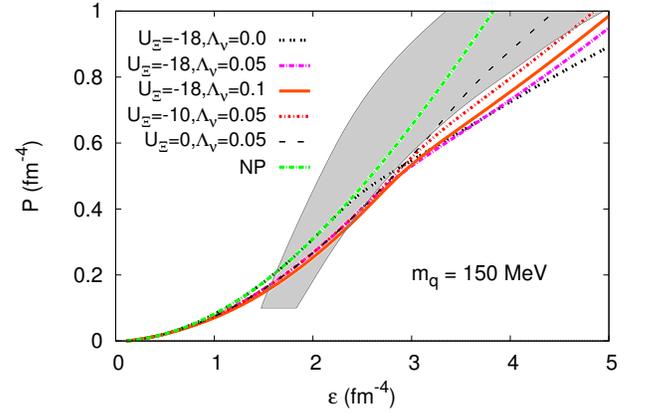}
\caption{(Color online) The EOS at various cascade potentials and different 
values of $\Lambda_{\nu}$ for quark mass $m_q=150$ MeV. The shaded region 
shows the empirical EOS obtained by Steiner {\it et al} from a 
heterogeneous set of seven neutron stars.}
\label{fig4}
\end{figure}
\begin{figure}
\includegraphics[width=8.cm,angle=0]{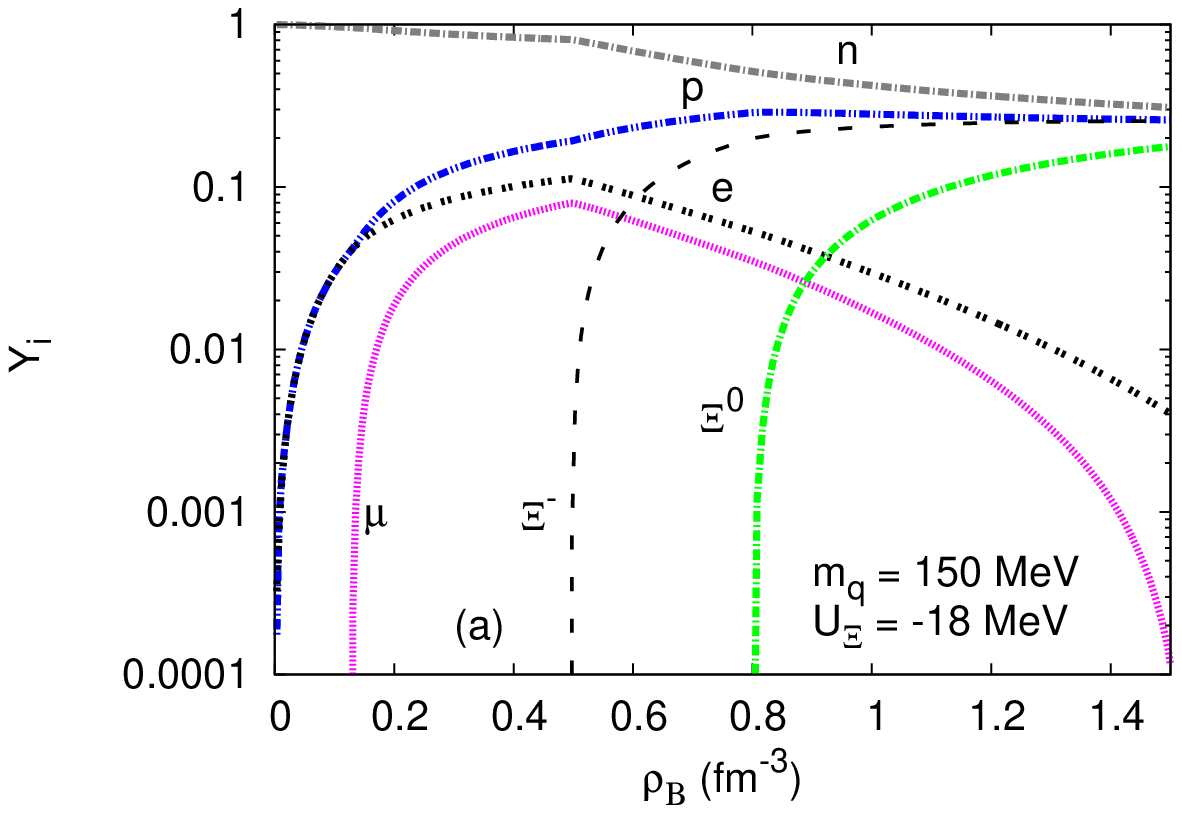}
\includegraphics[width=8.cm,angle=0]{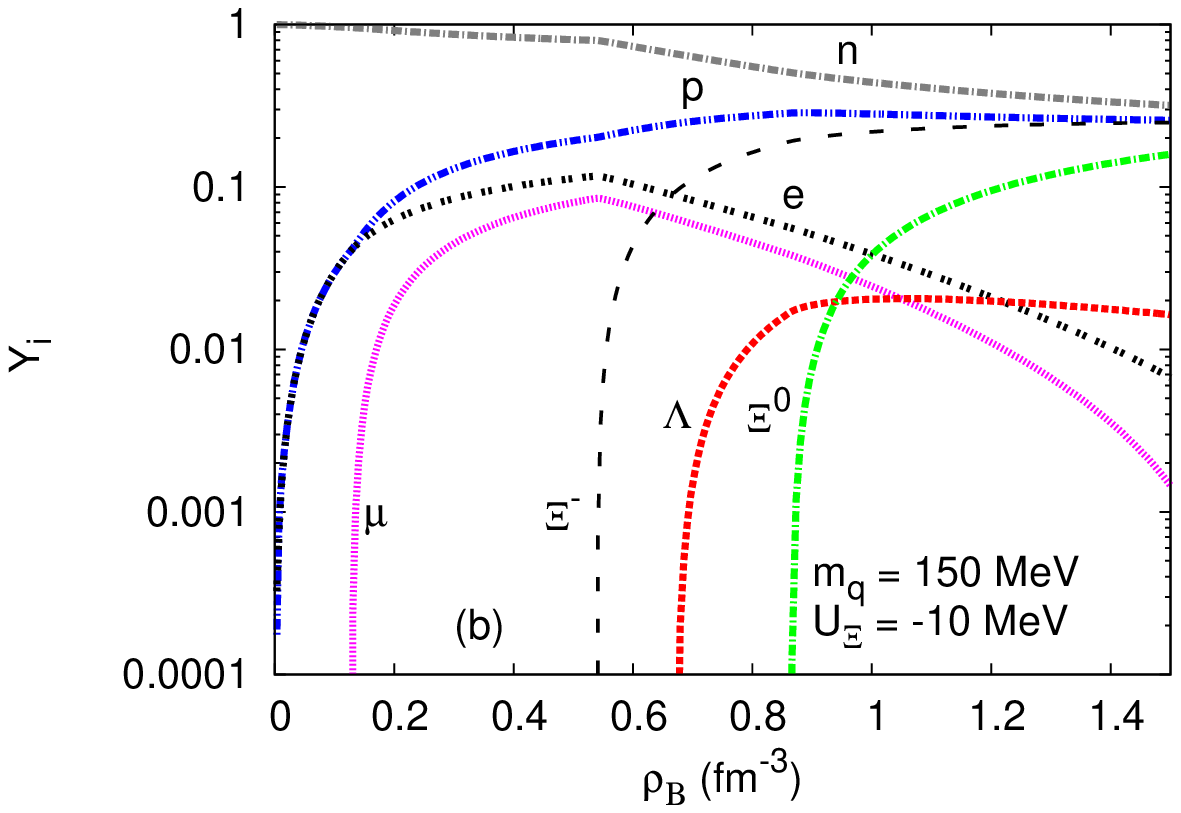}
\includegraphics[width=8.cm,angle=0]{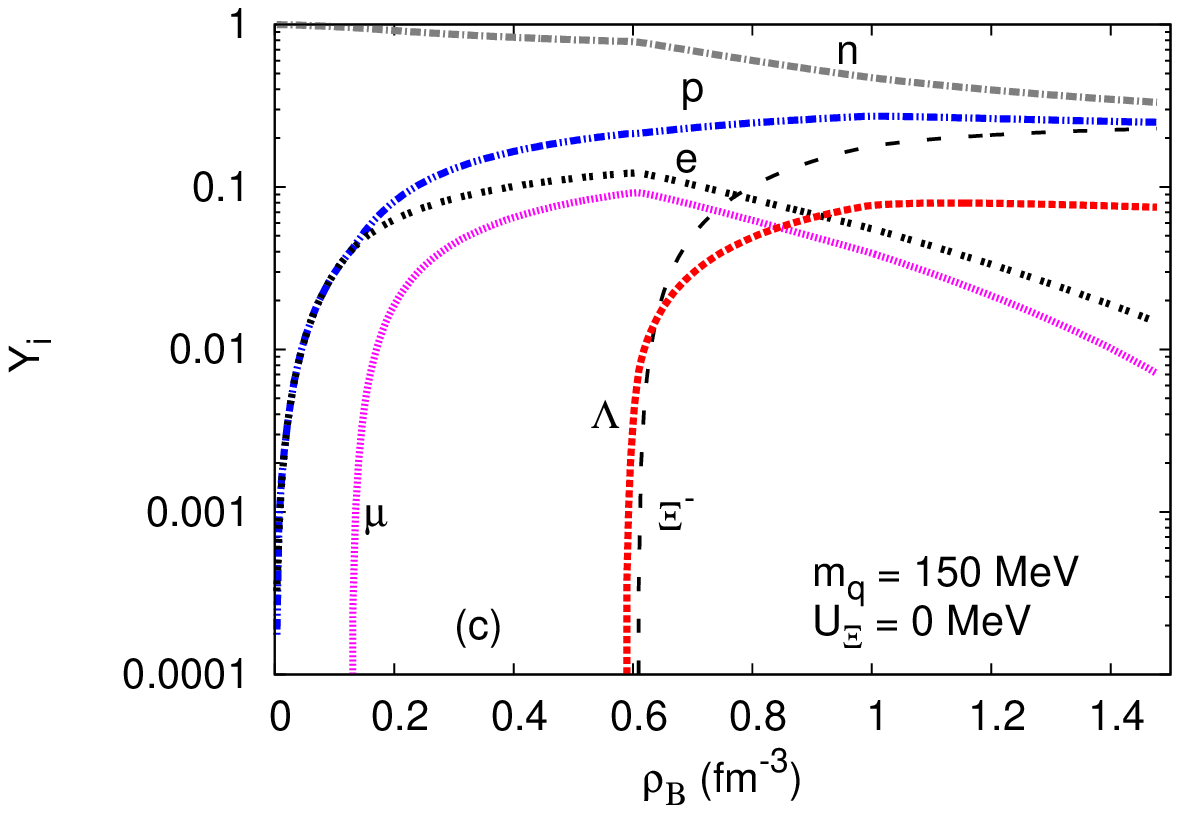}
\caption{(Color online) Particle fraction at different cascade potentials 
of (a) $U_\Xi=-18$~MeV, (b) $U_\Xi=-10$~MeV and (c) $U_\Xi=0$~MeV 
for the quark mass $m_q=150$ MeV.}
\label{fig5}
\end{figure}
The effect of including hyperons in neutron star matter is 
shown  in Fig. \ref{fig1}. It is observed that the EOS of neutron star matter 
with hyperons becomes softer starting from density $\rho_B=0.49$~fm$^{-3}$ 
compared to the one without the hyperons. The reason for such behaviour 
is that at  $\rho_B=0.49$~fm$^{-3}$ 
corresponding to $P\geq86.2$ MeV/fm$^{-3}$ or $P=0.437$~fm$^{-4}$ of neutron 
star matter, slow moving $\Lambda$, $\Sigma$ and $\Xi$ hyperons appear and 
the number of energetic nucleons and leptons decreases.

Fig. \ref{fig2} shows the effective baryon mass, $M_B^*/M_B$, as a function of 
baryon density. At saturation density $\rho_0$ the value of $M_B^*/M_B$ 
increases from $0.87$ for nucleons to $0.97$ for the $\Xi$ baryon. With 
increase in baryon density the effective mass decreases and then saturates at 
high baryon densities. 

The potentials that we have fixed for $\Lambda$, $\Sigma$ and $\Xi$ hyperons 
are plotted in 
Fig. \ref{fig3}. The hyperon potentials reduce with increasing density due 
to stronger repulsive effect at higher densities. In fact all hyperon 
potentials become repulsive nearly after twice the saturation density due to 
the non-linear density dependence of the baryon potentials.

In Fig. \ref{fig4} we plot the equation of state for quark 
mass $m_q=150$ MeV at different values of the coupling parameter 
$\Lambda_{\nu}$. The shaded region shows the empirical EOS obtained by 
Steiner {\it et al.} from a heterogeneous set of seven neutron stars with well 
determined distances \cite{awsteiner}.
We also show, for comparison, the EOS without the hyperons. 
The EOS with only neutron and proton (NP) matter is the stiffest and the 
corresponding star mass for quark mass $m_q=150$ MeV is $2.25$ $M_{\odot}$. 
The EOS with hyperons is softer than with NP matter. In fact, the softness 
increases by fixing the hyperon nuclear potentials at $U_\Lambda=-28$ MeV, 
$U_\Sigma=30$ MeV and $U_\Xi=-18$ MeV. Within such a set of potentials we 
observe that by increasing the coupling parameter $\Lambda_{\nu}$ the softness 
of the EOS increases with a corresponding decrease in radius. The effect of 
the variation 
in the values of the coupling parameter $\Lambda_\nu$ on the star mass and 
radius is given in Table \ref{table4}. 
\begin{table}[t]
\centering
\begin{tabular}{|c|c|c|c|c|c|c|}
\hline
$m_q$  &  $U_{\Xi}$  &  $\Lambda_{\nu}$  & $\varepsilon_0$ 
& M$_{max}$  &  R & R$_{1.4}$ \\
(MeV)  &  (MeV)  &   & ($fm^{-4}$)& ($M_{\odot}$)  & (km) & (km) \\ 
\hline
80                              &-18 &0    & 4.37 &1.81 &13.9 & 16.4  \\
                                &-18 &0.05 & 4.65 &1.70 &13.6 & 16.0 \\
                                &-18 &0.1  & 4.98 &1.64 &13.2 & 15.5 \\
                                &-10 &0    & 4.73 &1.85 &13.6 & 16.4 \\
                                & 0  &0    & 5.24 &1.88 &13.1 & 16.4 \\                     
\hline
150                             &-18 &0    & 3.52 &2.15 &15.6 & 19.2 \\
                                &-18 &0.05 & 3.99 &2.00 &14.9 & 19.1 \\
                                &-18 &0.1  & 4.38 &1.95 &14.4 & 18.8 \\
                                &-10 &0    & 3.75 &2.18 &15.2 & 19.2 \\
                                &-10 &0.05 & 4.28 &2.03 &14.6 & 19.1 \\
                                &-10 &0.1  & 4.66 &1.98 &14.1 & 18.9 \\
                                & 0  &0    & 4.03 &2.21 &14.9 & 19.2 \\
                                & 0  &0.05 & 4.64 &2.05 &14.1 & 18.9 \\
                                & 0  &0.1  & 5.07 &2.01 &13.7 & 18.8 \\
\hline
\end{tabular}
\caption{Stellar properties obtained at different values of the parameter
$\Lambda_{\nu}$ and the $\Xi$-meson coupling for quark mass $m_q=80$ 
MeV and $m_q=150$ MeV.}
  \label{table4}
\end{table}   
By changing $\Lambda_{\nu}$ from $0$ to $0.1$ 
the radius decreases by $\sim 1.2$ km and the mass of the star decreases 
by $0.2$ M$_\odot$. The variation in the softness with change in cascade 
potential $U_\Xi$ is studied. We observe that the EOS becomes stiffer for 
less attractive $U_\Xi$. Consequently we see that the mass increases by 
$0.06 M_\odot$ if $U_\Xi$ increases from 
$-18$ to $0$ MeV. This can be attributed to the fact that the hyperons 
occur at higher densities. For a comparison, we also show in 
Table \ref{table4} the radius corresponding to the canonical mass of 
$1.4$~M$_\odot$.

\begin{figure}
\includegraphics[width=10.cm,angle=0]{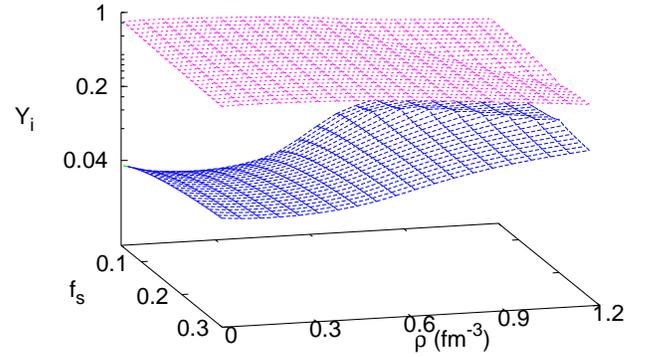}
\caption{(Color online) Strangeness fraction and 
particle fraction variation with density.The upper grid is for 
neutron and the lower one represents $\Xi^-$ hyperon. }
\label{fig6}
\end{figure}
Fig. \ref{fig5} shows the particle fractions for 
various fits of the cascade potential $U_{\Xi}$ in $\beta$-equilibriated 
matter. At densities below the saturation value the $\beta$-decay of 
neutrons to muons are allowed and thus muons start to populate. 
At higher densities the lepton fraction begins to fall since 
charge neutrality can now be maintained more economically with the 
appearance of negative hyperon species. In the present  case we observe the 
appearance of ${\Xi}^-$ first followed by $\Lambda$ baryon. Such a trend 
seems to be associated with our fittings of the cascade potential. At high 
densities all baryons tend to saturate. Given the growth of hyperons at higher 
densities, the dense interior of the star resembles more to a hyperon 
star than a neutron star.

Moreover, the $\Sigma$ hyperon is not present in the matter distribution 
for the given set of potentials since we have chosen a repulsive potential 
for it. The lepton fractions begin to drop at around $3\rho_0$. Hence to 
balance the positive charge of the protons the negatively charged $\Xi^-$ 
appear. It may be noted that the contribution of repulsive vector potential 
to the overall potential must be larger in order to prevent a collapse of the 
matter. The repulsive vector potential for $\Xi^-$ is smaller by a factor of 
two for other hyperons and by a factor of three for nucleons. In this light 
we can observe from Fig. \ref{fig5} that the $\Xi^-$ is more favoured to 
appear.

\begin{figure} 
\includegraphics[width=8.cm,angle=0]{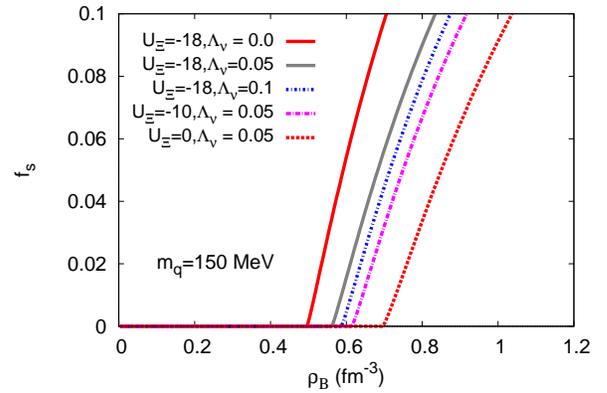}
\caption{(Color online) Strangeness fraction as
a function of density for various cascade potentials.}
\label{fig6b}
\end{figure}

The variation of the strangeness fraction and particle fraction of the 
$\Xi^-$ with density is compared to that of the neutron in Fig. \ref{fig6}.
The particle fraction of the $\Xi^-$ hyperon increases at higher densities 
implying the appearance of strangeness. With increasing densities the particle 
fraction of the neutron decreases and tends to saturate. We should note here 
that the strangeness content is sensitive to the meson-hyperon couplings. This 
can be observed in Fig. \ref{fig6b}. By increasing the cascade 
potential from $U_{\Xi}=-18$ MeV to $U_{\Xi}=0$ MeV, the onset of hyperons 
occurs at higher densitites. This makes the EOS stiffer for a less attractive 
potential.
\begin{figure*}
\includegraphics[width=8.cm,angle=0]{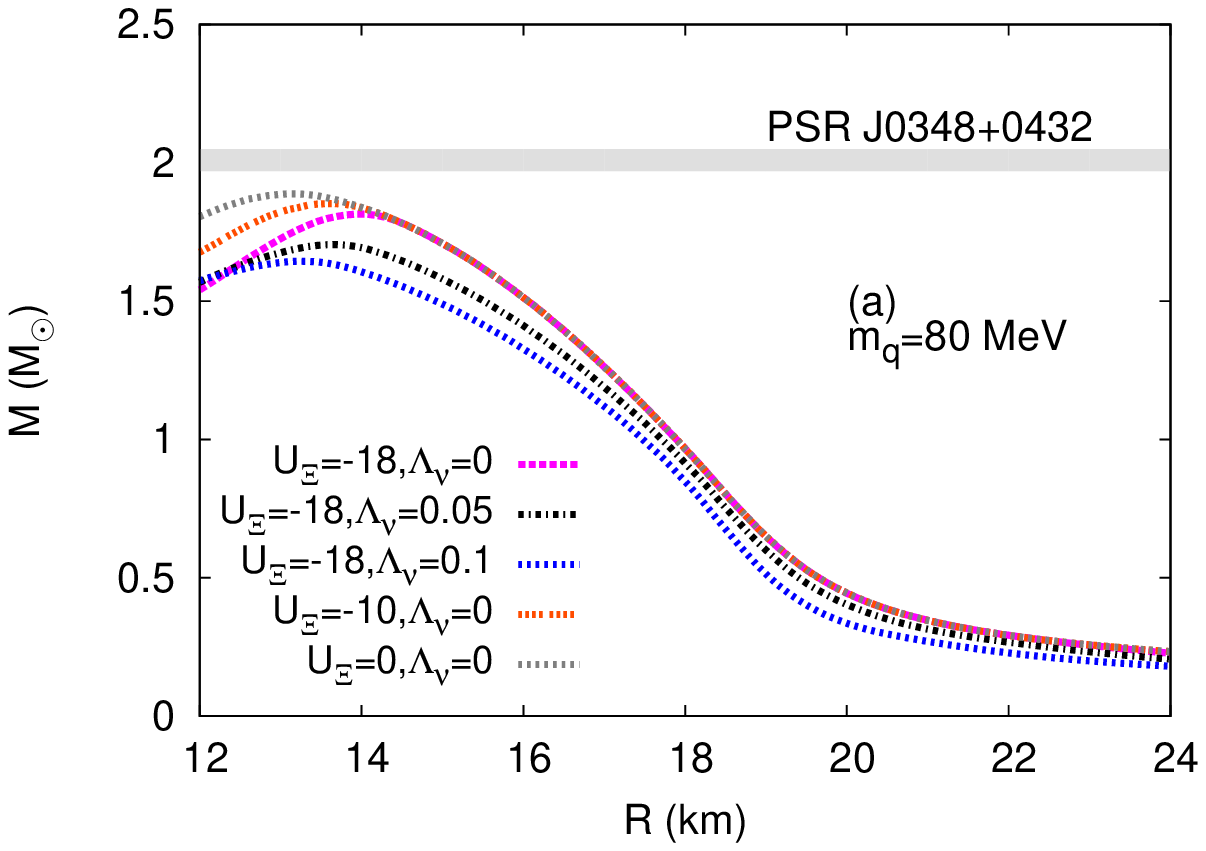}
\includegraphics[width=8.cm,angle=0]{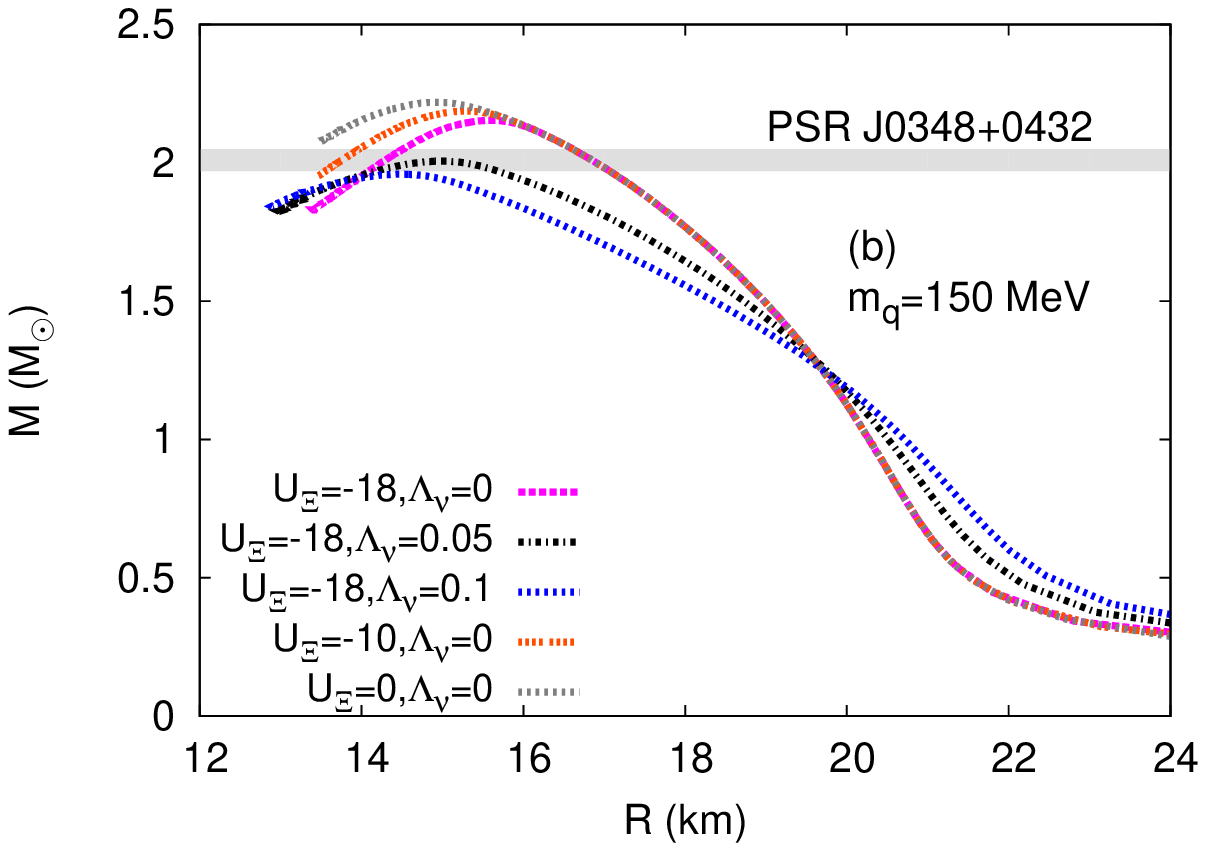}
\caption{(Color online) Star mass as a function of radius for various values 
of the coupling parameter and cascade potential at quark masses 
(a) $m_q=80$ MeV and (b) $m_q=150$ MeV. Also shown is the mass 
observed for the pulsar PSR J0348+0432 in \cite{antoniadis}.}
\label{fig7}
\end{figure*}

In Fig. \ref{fig7} we plot the mass-radius relations at two quark masses 
of $m_q=80$ MeV and $m_q=150$ MeV for the various 
scenarios and observe a direct correlation with the degree of 
stiffness of the EOS. As discussed earlier, for low values of the 
coupling parameter $\Lambda_{\nu}$ the EOS is stiffer giving higher mass 
as compared to higher values of $\Lambda_{\nu}$. Moreover, if we vary the 
cascade potential, we observe that for less attractive potential, the mass 
is the highest giving $M_{star}=1.88 M_{\odot}$ for quark mass $m_q=80$ MeV 
and $M_{star}=2.21 M_{\odot}$ at $m_q=150$ MeV.The detailed results are shown 
in Table \ref{table4}. The recently observed pulsar PSR J0348+0432 provide 
a mass constraint of $2.01\pm0.04 M_{\odot}$ \cite{antoniadis} while an 
earlier accurately measured pulsar PSR J1614-2230 gives a mass 
of $1.97\pm0.04 M_{\odot}$ \cite{demorest}. From our calculations we obtain a 
range of masses varying from $1.95 M_{\odot}$ to $2.21 M_{\odot}$ depending 
on the values of the coupling term as well as the variation of the cascade 
potential. The neutron star mass has been obtained under a similiar 
framework in QMC using bag model with variation 
in the values 
of $\Lambda_{\nu}$ and the cascade potential \cite{asysoft}. In this model 
the star mass obtained for the fixed 
cascade potential of $-18$ MeV gives a value $1.776$~M$_{\odot}$, 
$1.880$~M$_{\odot}$ and $1.888$~M$_{\odot}$ for $\Lambda_{\nu}=0,0.05$ and 
$0.1$ respectively. 

Though the measurement of the mass of the pulsars 
PSR J0348+0432 and PSR J1614-2230 is precise, the corresponding radii 
 measurements are not available. In fact, the simultaneous measurement of 
mass and radius of the same stellar object is uncertain. Radius measurements 
are primarily carried out from the studies of bursting neutron stars that show 
photospheric radius expansion \cite{lewin} and from transiently accreting 
neutron stars in quiscence \cite{rutledge}. Results from such measurements 
have been used to infer the pressure at several fiducial densities 
\cite{lindblom,latprak,readetal} 
as well as to 
put constraints on the neutron star equation of state at high densities 
\cite{oozel}. A recent study \cite{ozel} involving radius measurements to 
develop a neutron star equation of state predicts the radius to be 
$10.1$-$11.1$ km for a star of mass $M=1.5~M_{\odot}$. Another analysis 
\cite{awstein} encompassing variations in EOS and interpretations of the 
astrophysical data predicts the radius of a $M=1.4~M_{\odot}$ neutron star to 
lie between $10.4$ km and $12.9$ km. In the present work the radius  
corresponding to the canonical mass of $1.4~M_{\odot}$ is between $15.5-16.4$ 
km for $m_q=80$ MeV and between $18.8-19.2$ for $m_q=150$ MeV, which is 
quite higher than the radius range of $10.7$-$13.1$ km for 
$M=1.4~M_{\odot}$ stars \cite{newton,tsang,latlim} obtained from nuclear 
experimental studies. One of the reasons for this discrepancy on the 
radius may be due to the fact that the EOS considered here for the 
TOV equation does not include at high density, the effects of other 
phases of matter such as quark matter, mixed matter or paired quark matter.
However within the context of the present model, an 
improvement on this result may be explained by introducing additional 
interactions through $\delta$, $\sigma^*$ and $\phi$ meson exchanges 
without taking any other non-linear interactions.

\section{Conclusion}
In the present work we have developed the EOS using a modified quark-meson 
coupling model which considers the baryons to be composed of three 
independent relativistic quarks confined by an equal admixture of a 
scalar-vector harmonic potential in a background of scalar and vector mean 
fields. Corrections to the centre of mass motion, pionic and gluonic exchanges 
within the nucleon are calculated to obtain the effective mass of the baryon. 
The baryon-baryon interactions are realised by the quark coupling to the 
$\sigma$, $\omega$ and $\rho$ mesons through a mean field approximation. The 
nuclear matter incompressibility $K$ is determined to agree with experimental 
studies. Further, the slope of the nuclear symmetry is calculated which also 
agrees well with experimental observations.

The EOS is analyzed for different values of the non-linear coupling 
$\Lambda_{\nu}$ and quark mass. The variation in the degree of softness or 
stiffness of the 
EOS is concluded to be directly related to the higher or lower values of the 
coupling $\Lambda_{\nu}$ and quark mass $m_q$. 
The increase in the value of the coupling $\Lambda_{\nu}$ softens the 
EOS and decreases the maximum mass of star. In fact, we observe that there is 
no significant advantage of such a term in the context of obtaining the star 
mass constraint in the present set of parametrizations.

By increasing the quark mass the scalar coupling 
tends to be less sensitive to density variations, 
i.e., decreases more slowly and to fit to the nuclear matter 
properties more repulsion is required. 
The maximum star mass and 
strangeness fraction are 
quite sensitive to that, being larger/smaller by increasing/decreasing the 
quark mass.
Further, by fixing the hyperon-$\omega$ coupling from 
information of the hypernuclei as well as increasing the potential 
$U_{\Xi}$ to make it less attractive we have analyzed the variation in the 
stiffness of the EOS and the strangeness fraction at higher densities. We 
observe that the hyperon interactions influence the amount of strangeness 
in the star and thus have a strong impact on the maximum mass. We were able 
to obtain the observed mass of two accurately calculated pulsars, namely, 
PSR J0348+0432 and PSR J1614-2230 by varying the quark mass and 
cascade potentials, but more 
information on hypernuclei is required to further streamline the 
hyperon-meson couplings, such that we can constrain the quark mass parameter 
and strangeness fraction in the star. 

In the present set of parametrization, although we get the mass of the 
neutron star within the constraint of $2$~M$_\odot$, the radius corresponding 
to the canonical mass of $1.44$~M$_\odot$ is beyond the predicted values.   
From the studies of the effects of symmetry energy 
and strangeness content on neutron stars, Provid\^{e}ncia and 
Rabhi \cite{cprovindencia} observe that the radius of a hyperonic 
star of a given mass decreases linearly with the increase of the total 
hyperon content. By incorporating $\delta$, $\sigma^*$ and $\phi$ meson 
exchange contributions, we may expect some improvement in the 
prediction of the radius. Work in this direction is in progress.

\section*{ACKNOWLEDGMENTS}
The authors would like to acknowledge the financial assistance from 
BRNS, India for the Project No. 2013/37P/66/BRNS. TF thanks  the 
Brazilian agencies Coordena\c c\~ao de Aperfei\c coamento de Pessoal de N\'ivel Superior
(CAPES) and Conselho Nacional de Desenvolvimento Cient\'ifico e Tecnol\'ogico (CNPq) for partial support.

\end{document}